\documentclass[conference,compsoc]{IEEEtran}
\ifCLASSOPTIONcompsoc
  \usepackage[nocompress]{cite}
\else
  \usepackage{cite}
\fi
\ifCLASSINFOpdf
\else
\fi
\usepackage{graphicx}

\usepackage{multirow}
\usepackage{graphicx}

\begin{document}
%
\title{Having a Bad Day? Detecting the Impact of Atypical Life Events Using Wearable Sensors}


\author{\IEEEauthorblockN{Keith Burghardt\IEEEauthorrefmark{1},
Nazgol Tavabi\IEEEauthorrefmark{1},
Emilio Ferrara\IEEEauthorrefmark{2},
Shrikanth Narayanan\IEEEauthorrefmark{3},
Kristina Lerman\IEEEauthorrefmark{4}}
\IEEEauthorblockA{USC Information Sciences Institute}
}


%


\maketitle

\begin{abstract}
Life events can dramatically affect our psychological state and work performance. Stress, for example, has been linked to professional dissatisfaction, increased anxiety, and workplace burnout. We explore the impact of positive and negative life events on a number of psychological constructs through a multi-month longitudinal study of hospital and aerospace workers. Through causal inference, we demonstrate that positive life events increase positive affect, while negative  events increase stress, anxiety and negative affect. While most events have a transient effect on psychological states, major negative events, like illness or attending a funeral, can reduce positive affect for multiple days. Next, we assess whether these events can be detected through wearable sensors, which can cheaply and unobtrusively monitor health-related factors. We show that these sensors paired with embedding-based learning models can be used ``in the wild'' to capture atypical life events in hundreds of workers across both datasets. Overall our results suggest that automated interventions based on physiological sensing may be feasible to help workers regulate the negative effects of life events.

\end{abstract}


%
\IEEEpeerreviewmaketitle

\section{Introduction}

As organizations prepare their workforce for changing job demands, worker wellness has emerged as an important focus. Organizations see worker wellness as being central to their mission to develop a healthy and productive workforce while also maintaining optimal job performance. These goals are especially important in high-stakes jobs, such as healthcare providers working at hospitals, where job-related stress often leads to burnout and poor performance~\cite{gray1981stress,Bashir2010,Jamal2011}, and is one of the most costly modifiable health issues at the workplace \cite{Goetzel2012}. An additional challenge faced by workers is balancing demanding jobs with equally stressful events in their personal life. Adverse events---such as attending a funeral, the death of a pet, or illness of a family member---may amplify worker stress, and potentially harm job performance. On the other hand, positive life events---such as getting a raise, getting engaged, or taking a vacation---may decrease stress and improve well-being. The ability to detect such atypical life events in a workforce can help organizations better balance tasks to reduce stress, burnout, and absenteeism and improve job performance.

Until recently, detecting such life events automatically, in real time and at scale, would have been unthinkable. However, recent advances in sensing technologies have made wearable sensors more accurate and widely available, offering opportunities for unobtrusive and continuous acquisition of diverse physiological states.

Sensor-generated data, such as  heart rate and physical activity, allows for real-time, quantitative  assessment of individual's health~\cite{aral2017exercise} and psychological well-being~\cite{Healey2005,Hovsepian2015,Wang2014,Smets2018}
. Sensor data could also provide insights into atypical life events that individual workers experience and could affect their psychological well-being and job performance. However, the connection between atypical life events, individual well-being, and quantitative measurements from sensor data has not been demonstrated for such dynamic environments, especially in real-world scenarios.

In this paper, we report results of large longitudinal studies of hospital and aerospace industry workers who wore sensors and reported ecological momentary assessments (EMAs) over the course of several months. Workers also reported whether they had experienced an atypical event. The data allows us to use difference-in-difference analysis, a type of causal inference method \cite{Varian2016}, to measure the effect of atypical events---either positive or negative life events---on individual psychological states and well-being. We find that negative life events increase self-reported stress, anxiety, and negative affect by 10-20\% or more, while decreasing positive affect over multiple days. Positive life events, meanwhile, have little effect on stress, anxiety, and negative affect, but boost positive affect on the day of the event. Negative atypical events have a greater impact on  worker's psychological states than positive events, in line with previous findings~\cite{Baum2001}.

In addition to measuring the effects of atypical events, we show that it is possible to detect these events from a non-invasive wristband sensor. We discover that, although changes in individual psychological constructs are difficult to detect, atypical events are amenable to detection because they jointly affect several constructs. We propose a method that learns a representation of multi-modal physiological signals from sensors by embedding them in a lower-dimensional space. The embedding provides features for classifying when atypical events occur. Detection results are improved over baseline F1 scores by up to nine times, and achieve ROC-AUC of between 0.60-0.66. 

Physiological data from wearable sensors allows for studying individual response to atypical life events in the wild, creating opportunities for testing psychological theory about affect and experience. In addition, sensors data opens the possibility of passive monitoring to detect when individuals have stressful or negative experiences. While our initial results show that models can be further improved in the future, the ability to detect such experiences can help organizations improve the health and well-being of their workforce and reduce their detrimental effects on vulnerable populations.

\section{Related Work}

In this paper, we explore the effect of acute positive and negative events on human behavior, and how to detect these events with wearable sensors. 

We find that negative events increase stress, anxiety, and negative effect over the course of one to two days. Acute stress, in which stress increases over short periods \cite{Dimsdale2008}, can increase cardiovascular risk and depression \cite{Cohen2007}, and can negatively impact job performance \cite{gray1981stress,Bashir2010,Jamal2011}. 

Increased anxiety is associated with reduction in fertility \cite{Smeenk2001}, while negative affect is associated with higher sensitivity to pain \cite{Ruiz2011}. In this paper, we find that positive events increase positive affect. Higher positive affect is associated with broadened attention and improved creative problem solving \cite{Rowe2007,Lam2014}, and preferring future utility over present \cite{Ifcher2011}, although high levels may be associated with aversion to change \cite{Lam2014}. 

There exists extensive research on how sensors can be used to detect patterns and changes in human behavior 
\cite{Eagle2006,Wang2014}, including psychological constructs such as stress, anxiety, and affect (c.f., literature review of wearable sensors \cite{Banaee2013}). For example, they can detect if workers \cite{Srir2017} or students \cite{Sano2015} are stressed, even at a minute-by-minute level (c.f., cited literature in \cite{Can2019}).  Recent research has also explored detecting the degree to which a subject is stressed at shorter \cite{Smets2018,Healey2005,Can2019,Gjoreski2017,Sandulescu2015,Mozos2017}, and longer \cite{Guthrie1998,Edwards2010} timescales. Papers on stress typically induce stress externally \cite{Healey2005,Srir2017,Ghaderi2015,Camomilla2016}, but there are also papers on detecting natural stresses \cite{Can2019,Smets2018,Guthrie1998,Edwards2010}. While most related works have explored stress detection, there is some literature on detecting bio-markers associated with other psychological constructs. This includes anxiety \cite{Huang2017}, positive and negative affect \cite{Yan2019,Mottelson2016}, and depression \cite{Canzian2015}. In addition, recent literature has explored predicting (instead of detecting) multiple constructs using multi-task learning \cite{Jaques2017}. Notably, however, researchers needed access to data on social interactions, exercise, drug use, and sensors of several modalities, which may be unavailable in many situations. Finally, detecting acute positive and negative events is similar to research on using sensors for anomaly detection \cite{Banaee2013}. In contrast to previous literature, however, we detect events that affect psychological constructs rather than physiological constructs such as heart rate or sleep. 
In order to detect bio-marker patterns, sensors used in previous research measure a number of modalities including phone usage \cite{Sano2015}, skin conductance \cite{Healey2005,Villarejo2012,Srir2017,Smets2018}, heart rate \cite{Healey2005,Camomilla2016,Gjoreski2017,Srir2017,Smets2018,Hovsepian2015}, or breathing rate \cite{Healey2005} features.

The past work has suffered from two significant limitations. First, research has focused on either short time intervals (up to two weeks) and very small sample sizes (on the order of tens of subjects) \cite{Smets2018,Healey2005,Can2019,Gjoreski2017,Sandulescu2015,Mozos2017}, or collected data sporadically (once every several months) \cite{Edwards2010,Guthrie1998}. Second, previous literature has typically detected very short-term stresses (e.g., stresses that affect people on minute level \cite{Smets2018,Healey2005,Can2019,Gjoreski2017,Sandulescu2015,Mozos2017,Can2019}) rather than individual stressful events that impact someone over the longer term, such as funerals. Our work differs from these previous studies through continuous evaluation over several weeks of hundreds of subjects, allowing us to robustly uncover effects in diverse populations. Moreover, we uncover patterns associated with unusually good or bad events that can affect multiple psychological constructs over multiple days. 

\section{Data}

The data used in this paper comes from two studies aimed at understanding the relationship between individual variables, job performance, and wellness \cite{TILES2020}, which was part of the IARPA MOSAIC program. The study protocol was reviewed by USC Institutional Review Board (HS-17-00876 - TILES). Although the studies were conducted at different locations and recruited different populations, they had similar longitudinal design and collected similar data. The \textit{hospital} workforce data was collected during a 10-week long study that recruited 212 hospital workers. Participants were enrolled over the course of three ``study waves,'' each with different start dates (03/05/18, 04/09/18 and 05/05/18 for waves 1, 2 and 3 respectively). The \textit{aerospace} workforce data was collected from 264 subjects from 01/08/18 to 04/06/18. 

In both datasets, subjects' bio-behavioral data was captured via wearable devices. The studies also administered daily surveys to  collect self-assessments of individual participant stress, sleep, job performance, organizational behavior, and other personality constructs. The same survey questions were asked in both studies. We focus on positive affect, negative affect, anxiety and stress, which we discuss in greater detail in the psychological construct section.

In this paper, we use data collected from \textit{Fitbit} wristbands. Although other sensor data was collected during each study, including location data and audio or environmental features, we focus on this modality since it was common to both studies, and is the only sensor we have access to in the aerospace dataset. The Fitbit wristband captures dynamic heart rate and step count. It also offers a summary report of duration and quality of sleep for each day. Data is collected voluntarily by each subject, which was recorded at sub-minute levels. It was then uploaded to servers, where we aggregate the data. Table~\ref{tab:feature} summarizes the modalities captured by the Fitbit Charge 2 sensor. For the embedding approach, we only used the signals extracted from Fitbit (heart rate and steps) but for the aggregated method we also included the static summary features.



\begin{table*}[!t]
\centering
\begin{small}
\begin{tabular}{|c|l|}
\hline
\emph{Fitbit} & \multicolumn{1}{|c|}{\emph{Modality}} \\
\hline
\multirow{2}{11em}{\centering Signals (time series): 
}
& Heart rate (PPG) \\
& Number of steps \\
\hline
\multirow{5}{11em}{\centering Summary features (static): 
}
& Time in personalized heart rate ranges: ``fat burn,'' ``cardio,'' or ``out of range'' \\
& Daily minutes in bed \\
& Daily minutes asleep \\
& Daily sleep efficiency \\
& Sleep start \& end time \\
\hline
\end{tabular}
\end{small}\vspace{5pt}
\caption{Extracted features from sensors.}
\label{tab:feature}
\end{table*}

\begin{figure}
     \centering
     \includegraphics[width=0.48\columnwidth]{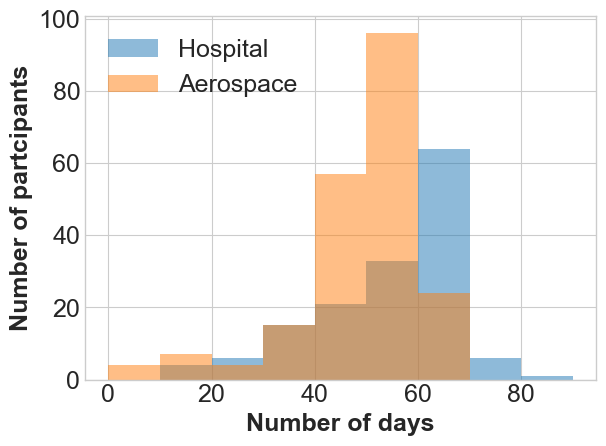}
     \includegraphics[width=0.48\columnwidth]{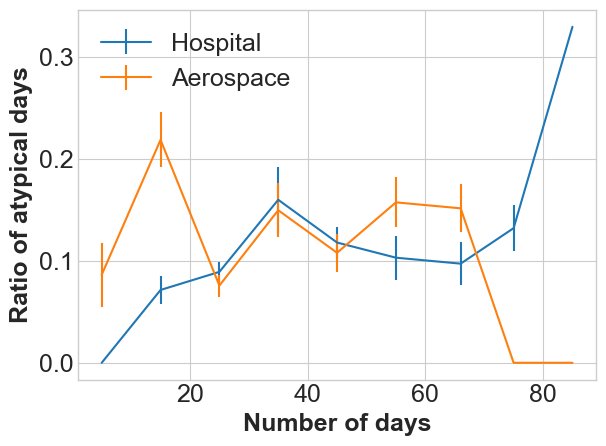}
     
     \caption{Statistics of compliance and frequency of atypical events. Left figure shows the number of days of data we have for each participant. Right figure shows the ratio of days in which there is an atypical event as a function of subject participation. Error bars are 95\% confidence intervals of the mean. 
     }
     \label{fig:hist}
 \end{figure}

Study participants exhibited varying compliance rates. As a result, collected data varied in the amount (hours per day) and length (number of days) across different participants. Figure \ref{fig:hist} shows the distribution of the data collected in these two datasets and the average ratio of atypical events for participants as a function of their compliance rate. We find in the left panel of Fig.~\ref{fig:hist} that most participants had several days of data, but a minority had only a few days of data over the entire study period. Pre-processing was therefore as follows. We only used data from participants who had at least  two days worth of data and one day marked as an atypical day. This brings the hospital data down to 8,155 days for 150 participants and the aerospace data to 10,057 days for 207 participants. We find in the right panel of Fig.~\ref{fig:hist} that removal of these low compliance subjects does not appear to significantly bias the data. 
Instead the frequency of atypical events is relatively independent of the compliance rate. 

The amount of data available from each day also varies and depends on the amount of time the participant wore the wristband. Although most participants (90\% in hospital dataset and 89\% in aerospace dataset) had the wristband on for the full day (24 hours), there are instances where only five hours of data could be collected in a day.

\subsection{Psychological Constructs}
The data used for this study includes daily self-assessments of psychological states provided by subjects over the course of the study. These  constructs include  self-assessments of job performance (Individual Task Proficiency (ITP) \cite{b17}, In-Role Behaviors \cite{b16}), Big Five personality traits (Openness, Conscientiousness, Extraversion, Agreeableness, Neuroticism~\cite{gosling2003very}), alcohol \cite{b24} and tobacco use~\cite{b25}, sleep quality \cite{buysse1989pittsburgh}, stress, anxiety, positive and negative affect. Stress and anxiety were measured by responses to questions that read, ``Overall, how would you rate your current level of stress?'' and ``Please select the response that shows how anxious you feel at the moment'' respectively and have a range of 1--5. Positive and negative affect were measured based on 10 questions from \cite{mackinnon1999short} (five questions for measuring positive affect and five for measuring negative affect) and have a range of 5--25. We focus on positive and negative affect, stress, and anxiety in this study because these were found to consistently change during an atypical event.

\subsection{Atypical event classification}
\label{sec:atypical-classes}
In addition to these constructs, subjects were also asked if they had experienced, or anticipated experiencing, an atypical event: ``Have any atypical events happened today or are expected to happen?''. If subjects replied yes, they had the option of add free-form text describing the atypical event. In the hospital data, there are  8,155 days of data, of which 958 days had atypical events (11.7\%). The aerospace data has 10,057 days of data, of which 1,503 were considered atypical (14.9\%). 

We have access to the free-form text in the hospital data, which was filled out by participants in 87\% of all atypical events. Surprisingly, the severity of the event could not be easily gleaned from sentiment analysis, such as VADER \cite{hutto2014vader} or LIWC \cite{Pennebaker2015}, as these tools gave neutral sentiment to text samples that were clearly negative. For example, text alike to ``at a funeral'' is given zero sentiment in VADER. We therefore applied a protocol, using human annotators, to categorize text as major negative events (such as deaths or injuries of loved ones), minor negative events (such as being stuck in traffic), or positive events (such as promotions). 
Major negative events were classified as negative life events such as major medical issues and funerals while minor negative events were daily hassles, sickness, or negative work events. Positive events were awards, promotions, weddings, and other events that were beneficial. Of all categorized atypical events, 210 (24\%) were positive, 626 (71\%) were minor negative events, and 39 (4.5\%) were major negative events.


\section{Methods}
\begin{figure*}[!t]
\centering
\includegraphics[width=0.95\linewidth]{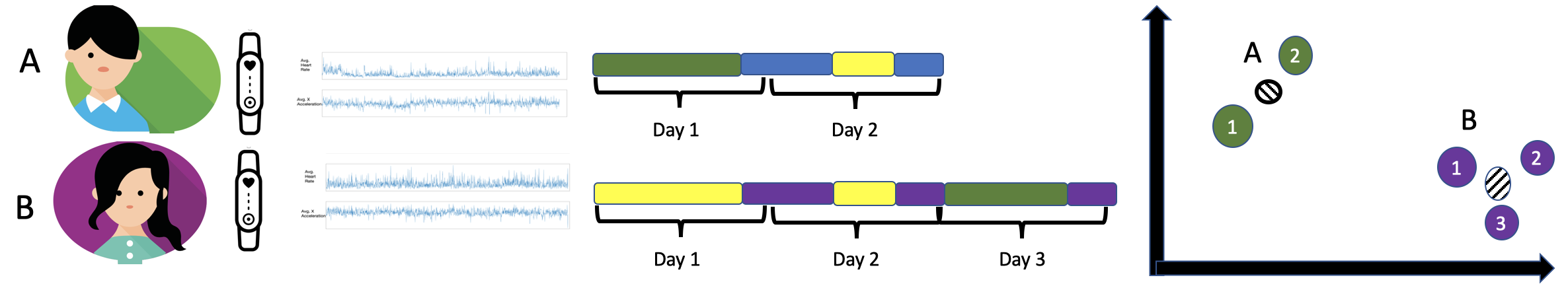}
\caption{Overview of the modeling framework. Sensor data collected from participants A and B (left two panels) is fed into non-parametric HMM model which outputs a state sequence per participant (middle panel), where states are shared among participants. Output from the HMM model is used to learn embeddings for each day of each participants (right panel).  The daily embedding (colored circles) and the average embedding for each participant (hashed circles) 
are used as features to detect an atypical day.}
\label{fig:framework}
\end{figure*}
\subsection{Causal Inference Method} 
The text descriptions of many atypical events in the hospital data mention sudden and unexpected events, such as 
an injured family member or unusually heavy traffic. We can therefore conjecture that atypical events create an as-if random assignment of any given subject over time. This is not always true, as in the case of subjects who report being on vacation multiple days, or are at different stages of burying a loved one. These are, however, relatively rare instances, with sequential events occurring in less than 15\% of atypical events in either dataset and exclusion of this data does not significantly affect results. To determine the effect of atypical events on subjects, we use a difference-in-difference approach to causal inference. Specifically, we look at all subjects who report an atypical event and then look at a subset who report stress, anxiety, negative affect, or positive affect the prior day. This is usually the majority of all events (83\%). We finally take the difference in their self-reported constructs from the day before the event. If subjects report construct values after the event (which is usually the case) we report the difference between these values and the day prior to an atypical event. We contrast these measurements with a \textit{null model}, in which we find subjects who did not report an atypical event on the same days that other subjects reported an atypical event, and find the change in their construct values from the prior day. This null model shows very little change in constructs over consecutive days, in agreement with expectation. The difference between construct values associated with the event and the null model is the \textit{average treatment effect} (ATE).

\subsection{Representation Learning}
\label{sec:learn}
We detect atypical events by embedding individuals' physiological time series data into a vector space, using the framework proposed in \cite{tavabi2019learning}. We then train models to identify where in this space do atypical events happen unexpectedly often. Namely, the time series is modeled as a hidden Markov model, where each state corresponds to an automatically inferred activity (e.g., exercising, working, or resting). The model effectively distinguishes activities people do during atypical days from activities during ``normal'' days.

In more detail, each subject's day of physiological data is interpreted as a multivariate time series, as described in Fig. \ref{fig:framework}, left two panels. The time series are transformed into sequences of hidden Markov states using a Beta Process Auto Regressive HMM (BP-AR-HMM) \cite{fox2014joint} (Fig. \ref{fig:framework}, center panel). Unlike classical hidden Markov models, BP-AR-HMM is flexible by allowing the number of hidden states to be inferred from the data. Based on these datasets the model found 73 states in the hospital data, and 130 states in the aerospace dataset, i.e., we find 73--130 ``activities'' that subjects perform, although they may only do a small fraction of these activities in a day. In addition, these states are shared among all subjects, rather than specific to one subject. This makes it feasible to embed data across different subjects and across different days. After the states are learned, we calculate the stationary distribution of time spent in each state to embed the daily data into the activity space (Fig. \ref{fig:framework}, right panel). This can be easily calculated from the HMM transition matrix by finding the eigenvector corresponding to the largest eigenvalue of the matrix.

\section{Results}

\begin{figure*}[tbh!]
    \centering
    \includegraphics[width=\textwidth]{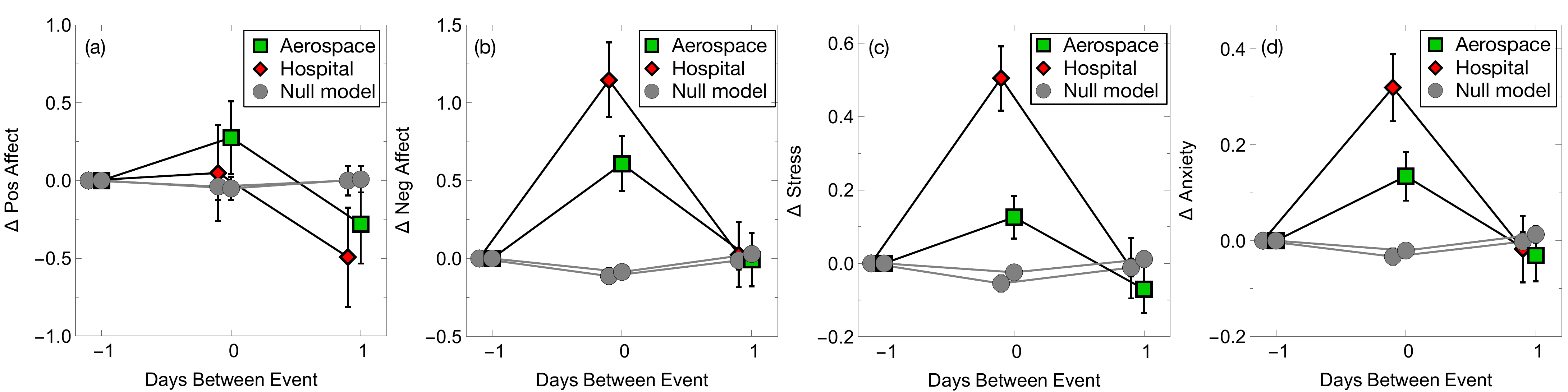}
    \caption{Effect of atypical events among the datasets studied. (a) Positive affect, (b) negative affect, (c) stress, and (d) anxiety. Green squares show the aerospace dataset, red diamonds show the hospital dataset, and gray circles are the null models, in which we collect sequential data from subjects who do not experience an atypical event at day zero.}
    \label{fig:diffdata}
\end{figure*}

How do atypical events affect individual's psychological states? We apply a difference-in-difference approach to measure the impact of atypical events on self-reported psychological constructs. We first look at the effect of atypical events across all our datasets, as shown in Fig.~\ref{fig:diffdata}. Atypical events, on average, have a relatively small effect on positive affect the day of the event (difference from null $= 0.09,~0.33$; p-value$=0.6,~0.009$, for hospital and aerospace data, respectively). We notice a decrease in positive affect from the day of the event to the day after the event 
(difference$=0.54,~0.55$; p-values $=0.0015,~0.017$ for aerospace and hospital data, respectively). On the other hand, there is a substantial increase in negative affect, stress, and anxiety (p-values $<0.001$), although changes are smaller in the aerospace dataset. 
\subsection{Causal Effect of Atypical Events}
\label{sec:caus}

\begin{figure*}[tbh!]
    \centering
    \includegraphics[width=\textwidth]{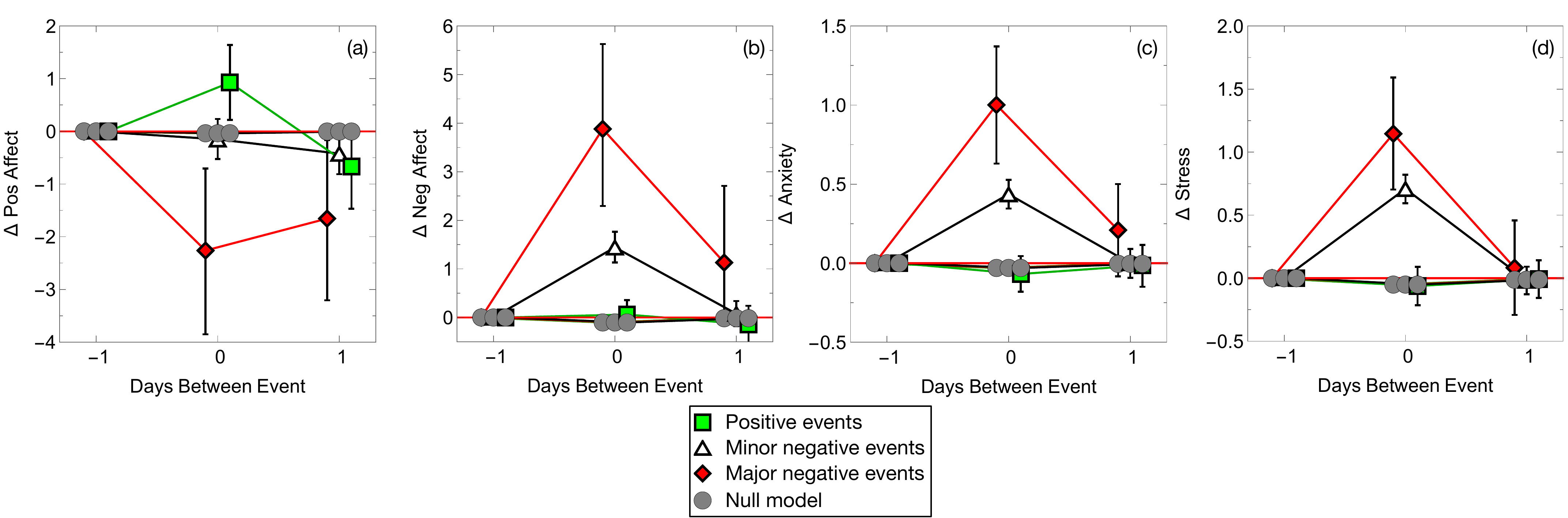}
    \caption{Effect of atypical events versus severity of event. (a) Positive affect, (b) negative affect, (c) stress, and (d) anxiety. Green squares are positive events, white triangles are minor negative events, red diamonds are major negative events, and gray circles are the null models. In the null models we collect sequential data from subjects who do not experience an atypical event at day zero.}
    \label{fig:diffevents}
\end{figure*}

The free-text descriptions that subjects provided about atypical events they experienced (only available in the hospital data), confirms these results. Most atypical events are negative, such as a fight with the spouse, traffic, or deaths. In a minority of cases, however, subjects report positive events, such as passing a test or a promotion. For the hospital data, we categorized atypical events as positive, minor negative, or major negative events, and determined the relative effect each has on subjects, as shown in Fig.~\ref{fig:diffevents}. We find that, as expected, positive events increase positive affect (p-value$=0.009$), but have no statistically significant effect on negative affect, stress, or anxiety (p-value $\ge 0.3$). Minor negative events do not substantially change positive affect on the day of the event (difference from null $=-0.15$, p-value$=0.57$), and have a small effect on positive affect the day after the event (difference from null $=-0.42$, p-value$=0.04$). On the other hand, they significantly increase negative affect, anxiety, and stress (p-value $<0.001$). Finally, major negative events both decrease positive affect the day of the event and the day after the event (p-value$=0.005,~0.03$ respectively). These results point to the strong diversity in atypical events, and support the idea that ``bad is stronger than good''~\cite{Baum2001}: adverse, or negative, events have a stronger effect on people than positive events, and are reported as atypical events more often.




\subsection{Detecting Atypical Events}

\subsubsection{Classification Task}
We evaluate performance of three classification tasks using sensor data: (1) detecting whether an atypical event occurred on that day; (2) detecting whether subjects experienced a good day; or (3) detecting whether subjects experienced a bad day. For (2) and (3) the classification task was ``1'' if subjects experienced a good or bad day, respectively, and ``0'' otherwise. Hence we simplify all tasks into a binary detection task. We emphasize that these last two tasks are only available for the hospital data. 

We use ten-fold cross validation. We choose to split datapoints at random, but in the Limitations section, we alternatively split users into training and testing sets to approximate a cold-start scenario where, in many cases, researchers train data on one cohort of subjects and classify data on another cohort \cite{Bogomolov2014}. The challenge of the latter detection task is that we need to classify if a subject has a good or bad day despite not training on any previous data from that subject. 
Performance metrics are averaged across all held-out folds. 

\subsubsection{Performance Metrics}
We use three performance metrics for evaluation. First, we use the area under the receiver operating characteristic (ROC-AUC) which quantifies how well a model can make true positives versus false positives. Random detection has an ROC-AUC of 0.5. Next, we use the F1 score, which is the harmonic mean of precision and recall. The higher F1 scores correspond to higher recall and precision of our estimates. Finally, we use precision itself as a performance metric because we want to determine the fraction of times we correctly label an atypical day (i.e., a ``good'' or ``bad'' day) as atypical. Low precision would indicate many false positives.

\begin{table*}[tbh!]
\centering
\begin{small}
\begin{tabular}{|c|c|c|c|c|c|}
\hline
{Dataset} & Construct & Model & \emph{ROC-AUC}&\emph{F1} & \emph{Precision} \\
\hline
\multirow{9}{6em}{\centering \textit{Hospital workforce}}&\multirow{3}{6em}{\centering Atypical Event} &Random& 0.50 & 0.12 & 0.12 \\
& &Aggregated&$0.57$ & $0.24$&$0.15$ \\
& &Embedding& $\mathbf{0.66}$ &$\mathbf{0.37}$ & $\mathbf{0.32}$  \\\cline{2-6}
&\multirow{3}{6em}{\centering Good Event} 
&Random& 0.50 & 0.03 & 0.03 \\
& &Aggregated&$\mathbf{0.63}$ & $0.08$&$0.04$ \\
& &Embedding& $0.62$ & $\mathbf{0.27}$ & $\mathbf{0.30}$\\
\cline{2-6}
&\multirow{3}{6em}{\centering Bad Event}
& Random& 0.50 & 0.08 & 0.08 \\
& &Aggregated&$ 0.57$ & $0.17$&$0.10$ \\
& & Embedding& $\mathbf{0.61}$ & $\mathbf{0.27}$ &$\mathbf{0.24}$\\
 \hline
\multirow{3}{6em}{\centering \textit{Aerospace workforce}} & \multirow{3}{6em}{\centering Atypical Event}& Random& 0.50 & 0.15 & 0.15 \\
& & Aggregated& $\mathbf{0.59}$ &$0.31$ & $0.21$\\
&  & Embedding& $\mathbf{0.60}$ & $\mathbf{0.32}$&$\mathbf{0.36}$ \\
\hline
\end{tabular}
\end{small}\vspace{5pt}
\caption{Performance of atypical event detection from sensors in the hospital and aerospace workforce datasets with randomly sampled cross-validation. For all datasets, we can classify whether an event is atypical. For hospital workers, we can also classify whether an event is ``good'' (instead of any other type of event), or ``bad.'' Percentages are above baseline (e.g., if classification is no better than random, the percentage would be 0\%).]
}
\label{tab:res}
\end{table*}

\subsubsection{Models}
We compare detection quality for two types of models: models using features from statistics of aggregated data, and models using features based on time series embeddings. 

{\bf Aggregated} We create several features based on aggregated statistics of signals and static modalities, listed in Table~\ref{tab:feature}. These statistics included the sum, mean, median, variance, kurtosis, and skewness of signals the day before, the day of, and the day after each day. Missing data is substituted with mean statistic value in the training or testing set. Statistics before and after each day were created because some physiological features, such as mean heart rate, might change before an atypical event, and some may change after, such as sleep duration. We use Minimum Redundancy Maximum Relevance on each dataset to select the best features (23 and 26 for the aerospace and hospital data respectively) \cite{MRMR}. Alternative features selection approaches using random forest feature importance produced poorer results. Typical features in the hospital data relate to sleep (for example, the top feature was tomorrow's minutes in bed). Typical features in the aerospace dataset tend to relate to heart rate (the top feature was the number of minutes in the ``fat burn'' heart rate zone in the past day). 

{\bf Embedding} when creating features from HMM embedding, we used only the signal modalities from Table~\ref{tab:feature}; the summary features were not used. Representations from HMMs were learned for the day of, and the day after each day. 
We also include the centroid of embeddings for each person in the training data as features, to control for subject-specific differences in behavior. We did not use any additional feature selection because embedding naturally reduces the feature dimensions. Imputation is also not needed because the HMM learns states based on the amount of data available for that day.

We use several candidate classification methods to detect whether a subject experiences an atypical event. For aggregate features, we compared logistic regression \cite{Cox1958}, random forest \cite{Ho1995}, support vector machines (SVMs) \cite{Cortes1995}, extra trees \cite{Geurts2006}, AdaBoost \cite{Freund1997}, and multi-layered perceptrons (MLPs) \cite{Hinton1989}. When training aggregate feature models, we make sure to downsample the majority class (no atypical event) such that the number of datapoints in each class are equal. Raw data, or upsampling the minority class, was found to produce worse results. Using all three performance metrics and ten-fold cross validation, we find atypical events in the hospital dataset are best modeled with random forests, while the aerospace workforce dataset is best modeled with logistic regression. In comparison, positive events are best modeled with random forests but negative events are best modeled with extra trees. 

Model hyperparameters for these models are chosen as follows. For random forest and extra trees, we used 100 trees and a max depth of 10. For AdaBoost, we let the number of estimators be 100. For all other hyperparameters, we use default parameters in Python library sklearn 0.21.3 for Python 3.7. For MLPs, we use three dense layers where the number of nodes in each layer equals the number of features in the model. For the model with embedding features, we used SVM, the same classifier used in the original work \cite{tavabi2019learning}. In all cases, hyperparameters were chosen as reasonable baselines, therefore additional improvements in model quality could be obtained with further tuning.

\subsubsection{Detection Results}

We demonstrate our model results in Table~\ref{tab:res}. First, we find that HMM embedding-based model outperforms alternative models. 
The ROC-AUC for the HMM-based model is 0.60 for the aerospace workforce and 0.66 for the hospital workforce. Positive and negative events similarly have an ROC-AUC of 0.61-0.63. F1 and precision exceed random baselines by factors of two to nine. The seemingly low F1 and precision are due to the rarity of atypical events, especially for positive events, which only happen on 3\% of days, and negative events which only happen in 8\% of all days. A detection therefore represents a ``warning sign'' that a worker may have had an negative event that day. Overall, detecting atypical events shows promise.

\section{Discussion}

Our results highlight how unusual but impactful events strongly affect workers. Interestingly, however, atypical events are more often negative than positive. For example, 8\% of all days among hospital workers contained negative events, while only 3\% contained positive events. The relative adversity and frequency of negative events over positive events in our data agrees with previous findings that negative events are often more impactful \cite{Baum2001}. Moreover, we find that significant events cannot be viewed as affecting a single psychological construct; they can affect multiple constructs at once. In the same way that multi-task learning can improve predictions in AI \cite{Ruder2017}, we expect that atypical event detection could be useful to detecting anxiety, stress, and other psychological constructs simultaneously.

Our results also point to important future work. First, while the performance of our method does not allow it to be used in practice, it can be considered a significant starting point. Other sensor modalities can be added to better infer when or if an atypical event occurs. These include breathing, skin conductance, or phone usage sensors. Next, personalizing our methods to individuals has the potential to substantially improve detection performance \cite{Jaques2017}. We find, for example, some subjects experience very few atypical events while others experience atypical events triple the average rate. Next, we can extend our results by analyzing how similar good or bad events affect people differently. Some subjects may be able to cope with negative events better than others. 
\begin{table*}[!tbh]
\centering
\begin{small}
\begin{tabular}{|c|c|c|c|c|c|}
\hline
\emph{Dataset} & Construct & Model & \emph{ROC-AUC}&\emph{F1} & \emph{Precision} \\
\hline
\multirow{9}{6em}{\centering \textit{Hospital workforce}}&\multirow{3}{6em}{\centering Atypical Event} &Random& 0.50 & 0.12 & 0.12 \\
& &Aggregated&$0.55$ & $0.22$&$0.14$ \\
& &Embedding& $\mathbf{0.56}$ &$\mathbf{0.23}$ & $\mathbf{0.16}$  \\\cline{2-6}
&\multirow{3}{6em}{\centering Good Event} 
&Random& 0.50 & 0.03 & 0.03 \\
& &Aggregated&$0.57$ & $0.065$&$0.035$ \\
& &Embedding& $\mathbf{0.58}$ &$\mathbf{0.08}$ & $\mathbf{0.05}$ \\
\cline{2-6}
&\multirow{3}{6em}{\centering Bad Event}
& Random& 0.50 & 0.08 & 0.08 \\
& &Aggregated&$\mathbf{0.57}$ & $0.15$&$0.09$ \\
& & Embedding& $0.56$ & $\mathbf{0.16}$ &$\mathbf{0.10}$\\
 \hline
\multirow{3}{6em}{\centering \textit{Aerospace workforce}} & \multirow{3}{6em}{\centering Atypical Event}& Random& 0.50 & 0.15 & 0.15 \\
& & Aggregated& $\mathbf{0.58}$ &$\mathbf{0.30}$ & $\mathbf{0.20}$\\
&  & Embedding& 0.54 & $0.25$&$0.17$ \\
\hline
\end{tabular}
\end{small}\vspace{5pt}
\caption{Performance of atypical event detection from sensors in the hospital and aerospace workforce datasets with user held-out detection. For all datasets, we can classify whether an event is atypical. For hospital workers, we can also classify whether an event is ``good'' (instead of any other type of event), or ``bad.'' 
}
\label{tab:res_uid}
\end{table*}

\section{Limitations}


There are, however, a number of limitations we should discuss, that highlight limitations in the data, as well as broader model limitations that offer implications for model design.

First, data was only collected once a day, and we were unable to gather when atypical events occurred during the day. This made the detection problem much harder because there are a number of separate reasons for heart rates or step counts to change and inferring the specific signal that would indicate an atypical event is unavailable in our data. Next, we are limited in the modalities we had access to, and therefore the physiological behavior we could measure. For example, stress might be more accurately measured with the help of skin monitors \cite{Healey2005,Villarejo2012,Srir2017,Smets2018}. 

Finally, our results are based on cross validation, a standard method in which datapoints are divided into training and testing splits. This is alike to previous work on detecting stress, in which training and testing was performed on the same users \cite{Healey2005,Can2019,Sandulescu2015,Mozos2017}. It's feasible, however, that a model may be trained on one dataset and tested on another. To approximate this scenario, we instead split users, rather than days of data, into training and testing folds. We show our model performance results in Table~\ref{tab:res_uid}.  Atypical events can be detected 91--220\% above baselines based on F1 score, but results are more modest than in the Results section, with a reduction in ROC-AUC from 0.66 to 0.58 for hospital atypical events. These results are alike to other recent papers, which split subjects into training and testing and found relatively poor model performance \cite{Smets2018,Gjoreski2017}. On one hand, this means that these models will not necessarily be able to work out of the box. They need to be personalized to users. That said, once they are tuned to the cohort, the performance is respectable. Human heterogeneity therefore make physiologically-based psychological modeling especially difficult.

\section{Conclusion}
We discover that atypical events and negative events substantially increase stress, anxiety, and negative affect. Major negative events are found to reduce positive affect over multiple days, while positive events improve positive affect that day. We also demonstrate that wearable sensors can provide important clues about whether someone is experiencing a positive or negative event. We find atypical events can be predicted with ROC-AUC of 0.66 with relatively little model hyperparameter tuning. This suggests more improvements are possible to predict atypical events. Overall, these results point to the importance and relative detectability of negative events, which offer hope for remote sensing and automated interventions in the future.



\ifCLASSOPTIONcompsoc
  \section*{Acknowledgments}
\else
  \section*{Acknowledgment}
\fi

The authors are grateful to the \textit{TILES} team for the efforts in study design, data collection and sharing that enable this work. 
This research is based upon work supported by the Office of the Director of National Intelligence (ODNI), Intelligence Advanced Research Projects Activity (IARPA), via IARPA Contract No 2017-17042800005.



%



\end{document}